\documentclass[
 reprint,
nofootinbib,
 amsmath,amssymb,
 aps,
prb,
floatfix,]{revtex4-2}
\usepackage{bm}
\usepackage{latexsym}
\usepackage{dcolumn}
\usepackage{amsmath,amsfonts,amssymb}
\usepackage{graphicx,epsfig}
\usepackage{psfrag}
\usepackage{amsmath,amssymb}
\usepackage{amsthm}
\usepackage{color}
\usepackage{wasysym,breqn}
\usepackage{natbib}
\usepackage{moresize}
\usepackage{enumerate}
\usepackage[hang,flushmargin]{footmisc}
\usepackage[titletoc,toc]{appendix}
\usepackage{lipsum}
\usepackage{subcaption}
\usepackage{epstopdf} 
\usepackage[pdfencoding=auto]{hyperref}
\usepackage{bookmark}
\usepackage{rotating}
\usepackage{multirow}
\usepackage{mathtools}
\usepackage[font=small,labelfont=bf,
   justification=justified,
   format=plain]{caption}

\usepackage{stackengine}
\usepackage{scalerel}
\usepackage{tabularray}
\usepackage{xtab,afterpage}
\newcommand{\be}{\begin{equation}}
\newcommand{\ee}{\end{equation}}
\newcommand{\bea}{\begin{eqnarray}}
\newcommand{\eea}{\end{eqnarray}}
\newcommand{\bse}{\begin{subequations}}
\newcommand{\ese}{\end{subequations}}
\newcommand{\bce}{\begin{center}}
\newcommand{\ece}{\end{center}}
\newcommand{\bfg}{\begin{figure}}
\newcommand{\efg}{\end{figure}}
\newcommand{\bit}{\begin{itemize}}
\newcommand{\eit}{\end{itemize}}
\newcommand{\bed}{\begin{description}}
\newcommand{\eed}{\end{description}}
\newcommand{\ben}{\begin{enumerate}}
\newcommand{\een}{\end{enumerate}}

\newcommand{\bdm}{\begin{displaymath}}
\newcommand{\edm}{\end{displaymath}}

\begin{document}


\title{Cosmological Singularity and Power-Law Solutions in Modified Gravity}

\author{Saurya Das}
 
 \email{Email:saurya.das@uleth.ca}
\affiliation{Department of Physics and Astronomy \& Quantum Alberta \\
University of Lethbridge\\
4401 University Drive \\Lethbridge AB T1K 3M4 Canada }

\author{S. Shajidul Haque}
 \email{Email:shajid.haque@uct.ac.za}
\affiliation{Department of Mathematics and Applied Mathematics\\
University of Cape Town\\ Rondebosch 7700 Cape Town South Africa\\ and \\National Institute for Theoretical and Computational Sciences (NITheCS)\\Private Bag X1 Matieland\\South Africa}

\author{Seturumane Tema}
 \email{Email:tmxset$001$@myuct.ac.za}
\affiliation{Department of Mathematics and Applied Mathematics\\University of Cape Town\\ Rondebosch 7700 Cape Town South Africa}
\begin{abstract}
A bouncing Universe avoids the big-bang singularity. 
Using the time-like and null Raychaudhuri equations, 
we explore whether the bounce near the big-bang, within a broad spectrum of modified theories of gravity, allows for cosmologically relevant power-law solutions under reasonable physical conditions. Our study shows that certain modified theories of gravity, such as Stelle gravity, do not demonstrate singularity resolution under any reasonable conditions, while others including $f(R)$ gravity and Brans-Dicke theory can demonstrate 
singularity resolution under suitable conditions. 
For these theories, we show that the accelerating 
solution is slightly favoured over ekypyrosis.
\end{abstract}
\maketitle

\section{Introduction}
It has been established beyond reasonable doubt that our
Universe is expanding and accelerating, {\it provided} one assumes the validity of the General Theory of Relativity (GR). 
While this is a reasonable assumption given the success of GR, such as in the planetary, galactic and cosmological regimes, providing us with a rather simple model
of our Universe, which is spatially flat and has a significant amount of Dark Matter and Dark Energy at the present epoch, it is also a cause of concern for the following reasons. First, one is far from having a definitive answer about the true nature of Dark Matter and Dark Energy or even if Dark Energy is really there, 
and second, extrapolating backwards in  time, it is easy to show that the scale factor of the Universe, $a(t)$, vanishes and curvature scalars blow-up about $13.7$ Gyr in the past, resulting in the putative big-bang singularity. This implies that (a) about $95\%$ of the Universe, the sum of the proportions of Dark Matter and Dark Energy in the present epoch, is made up of unknown constituents, and (b) the fundamental laws of physics as we know, together with their predictive powers, break-down at the singularity. 

{To address the singularity issue, we initially explore modified theories of gravity, evaluating their viability for power-law solutions. Specifically, we examine ekpyrosis \cite{khoury2001ekpyrotic,steinhardt2002cyclic} and accelerated power-law solutions, representing alternative cosmological scenarios to inflation and a cosmological constant, respectively. These solutions are tested for both the strong energy and null energy conditions via the Raychaudhuri equation.

More precisely, we investigate three concrete models of modified gravity and demonstrate that, in two out of three cases, if the singularity is to be avoided through a bounce—meaning the scale factor never reaches zero but enters an expanding phase very close to the singularity—we can attain both types of power-law solutions. However, the conditions slightly favor accelerating power-law solutions over ekpyrosis. Specifically, a bouncing Universe via modified gravity tends to align more with the accelerating solution. If this trend holds generally, it would offer a unification of two research areas aimed at resolving the big-bang singularity. 
}


To study the existence of singularities in general and in particular at the big-bang, we use the Raychaudhuri equation, which as we know, governs the evolution of `expansion of geodesics', and is the primary tool behind the celebrated Hawking-Penrose singularity theorems \cite{hawking1970singularities, penrose1965gravitational,hawking2023large,hawking2010nature, ellis1977singular}. The equation predicts the existence of conjugate points, both in the finite past and the finite future. A congruence of geodesics meet at these points, making the tangent vectors ambiguous. This in turn implies the incompleteness of geodesics via a set of straightforward arguments, which is generally accepted as the necessary and sufficient condition for a spacetime to be singular \cite{senovilla1998singularity}. The only assumption that one needs to make is that the energy-momentum under question satisfies reasonable energy conditions. These can of course be broken under certain circumstances, for example when one has a positive cosmological constant (which has negative pressure) or when certain modified theories of gravity are considered instead of GR \cite{nojiri2011unified,nojiri2017modified}. 

Our paper is organized as follows. In the next section, we do a quick review of the power-law solutions in cosmology. 
In Section III, we review the Raychaudhuri equation and its prediction of conjugate points and singularities under very general circumstances. 

In Sections IV, V and VI, we study the corrections to the cosmological backgrounds and bounces in the context of {Stelle-gravity}, $f(R)$ and {Brans-Dicke}  theory, arguably some of the best studied and well-motivated modifications of GR. We collect our results, summarize and discuss future directions in Section VII.

\section{Cosmology and Power-law Solutions}
Our objective is to examine whether various modified gravity models result in cosmological power-law solutions, 
when avoiding the singularity. These solutions serve as cosmological attractors \cite{ali2014power}. For example, an exponential potential yields a scale factor with a power-law behavior. Our investigation covers both late-time accelerating solutions and ekpyrotic solutions for the early universe. For power-law solutions, the scale factor is as follows:
\begin{equation}
a(t) \sim t^{\alpha}
\end{equation}
We get accelerated expansion when 
$\alpha>1$ and ekpyrosis when 
$\alpha<1/3$. This paper specifically concentrates on these two types of solutions. Other values of $\alpha$ correspond to generic power-law solutions, which are of no direct interest to our current work. 

Inflation proposes that the universe underwent significant acceleration during its early stages, rapidly expanding from subatomic to macroscopic scales within a fraction of a second \cite{guth1981inflationary,linde1982new}. However, the adoption of accelerating power-law solutions as a model for inflation is disregarded due to their large tensor-to-scalar ratio, as initially identified in \cite{Souradeep:1992sm}, which conflicts with the Planck data. Consequently, the accelerating scenario becomes a feasible option solely for late-time cosmology.

Ekpyrotic or cyclic cosmology emerges as a popular alternative to the theory of cosmic inflation. It posits that our universe undergoes an infinite number of cycles, each beginning with a big bang phase. This is followed by a slow accelerating expansion phase, then a slow contraction phase, ultimately culminating into a big crunch. The power-law behavior becomes evident during the slow contraction phase of the ekpyrotic universe's evolution
\cite{khoury2001ekpyrotic,steinhardt2002cyclic}. 
Therefore, searching for ekpyrotic solutions is mathematically very similar to the search for accelerating power-law solutions. 

\section{Raychaudhuri equation and the focusing Theorem}

In this section, we study the bounce in the 
language of the Raychaudhuri equation, which succinctly displays the existence of caustics or conjugate points for a congruence of time-like or null geodesics for most reasonable spacetimes and corresponding stress-energy tensors, which is essential for the singularity theorems to hold. Consequently, 
conditions which may prevent the formation of such caustics is a useful tool in studying singularity resolution. In particular, one is looking for additional terms in the RHS of the Raychaudhuri equation, which are positive, as these are the ones which can potentially prevent the formation of caustics, and thereby render the singularity theorems inapplicable.

Geodesic congruences with time-like and null tangent vector fields are characterized by their
expansion $\theta = \nabla_{\mu} u^{\mu}$, 
$\hat \theta= \nabla_{\nu} n^{\nu} $, respectively\cite{haque2017consistent}, which satisfy the Raychaudhuri equations
\begin{eqnarray}\label{lotu}
\frac{d\theta}{d\tau} &=& -\frac{\theta^2}{D-1} - R_{\mu\nu}u^{\mu}u^{\nu} \ \ \rm(time-like)\cr 
\frac{d\hat{\theta}}{d\lambda} &=&-\frac{\hat{\theta^2}}{D-2}-R_{\mu\nu}n^{\mu}n^{\nu} \ \ \rm(null) \,
\end{eqnarray}
where $D$ is the number of spacetime dimensions and $\tau$ and $\lambda$ are the proper time and affine parameter respectively. 
In General Relativity one can trace-reverse the Einstein field equations to get
\begin{equation}\label{khf}
R_{\mu\nu}= 8\pi G\left(T_{\mu\nu}-\frac{1}{2}Tg_{\mu\nu}\right)
\end{equation}
where $T=g^{\mu\nu}T_{\mu\nu}$ is the trace of the energy-momentum tensor $T_{\mu\nu}$. Contracting equation (\ref{khf}) with the time-like vectors $u^{\mu}$ results in
\begin{equation}
R_{\mu\nu}u^{\mu}u^{\nu} = 8\pi G\left(T_{\mu\nu}-\frac{1}{2}Tg_{\mu\nu}\right)u^{\mu}u^{\nu}\,
\end{equation}
which further simplifies to
\begin{equation}\label{gobe}
R_{\mu\nu}u^{\mu}u^{\nu} = 4\pi G \left(\rho+3p\right)
\end{equation}
for a perfect fluid with energy-momentum tensor of the form: $T_{\mu\nu}=\left(\rho+p\right) u_{\mu}u_{\nu}+p g_{\mu\nu}$. 

When $d\theta/d\tau \leq0$ the time-like geodesics converge to a focal point \cite{raychaudhuri1955relativistic,hawking1970singularities,ellis2003closed, penrose1965gravitational,hawking2023large,das2019constraints}. Since $\theta^2$ is positive, for $D=4$ the convergence condition simplifies to proving only
\begin{equation}\label{conv}
R_{\mu\nu}u^{\mu}u^{\nu} \ge 0
\end{equation}
When matter satisfy the strong energy condition 
\begin{equation}\label{nullcond}
\rho+3p \ge0\,
\end{equation} 
the above inequality is immediately satisfied. Note that, for a perfect fluid with the equation of state $p=w\rho$ the strong-energy condition reads:
\begin{equation}\label{weakener}
w\ge -\frac{1}{3}.
\end{equation}
Therefore, ordinary matter such as radiation ($w=\frac{1}{3}$) and dust ($w=0$) obey (\ref{weakener}), however, a positive cosmological constant $\Lambda$ ($w=-1$) is a classic example that violates the strong energy condition.

Now, when we contract (\ref{khf}) with the null vectors $n^{\mu}$ we get
\begin{equation}
R_{\mu\nu}n^{\mu}n^{\nu} = 8\pi G T_{\mu\nu}n^{\mu}n^{\nu}=8\pi Ga^{-2}\left(\rho+p\right).
\end{equation}
Therefore, the convergence condition for null geodesics reduces to showing
\begin{equation}\label{qer}
\rho+p\geq 0\,
\end{equation}
which is referred to as the Null energy condition. Note that for a perfect fluid equation (\ref{qer}) translates to
\begin{equation}\label{azi}
w\geq-1 .
\end{equation}

In this paper, we consider power-law solutions for both cases—the strong-energy condition (\ref{weakener}) and the null energy condition separately. However, we require that the convergence condition (\ref{conv}) is violated. This occurs when there exists a positive term on the right-hand side of both equations in (\ref{lotu}), originating from the $R_{\mu\nu} u^{\mu} u^{\nu}$ ($R_{\mu\nu} n^{\mu} n^{\nu}$) of the modified gravity theory, which dominates over the first term $\frac{\theta^2}{D-1}$ $\left(\frac{\hat{\theta}^2}{D-2}\right)$ respectively for the time-like (null) Raychaudhuri equations.
\begin{eqnarray} \label{violation}
    \frac{d\theta}{d\tau} &=& -\frac{\theta^2}{D-1} - R_{\mu\nu}u^{\mu}u^{\nu} >0 \ \ \rm (time-like)\cr
\frac{d\hat{\theta}}{d\lambda} &=&-\frac{\hat{\theta^2}}{D-2}-R_{\mu\nu}n^{\mu}n^{\nu} >0 \ \ \rm (null)
\end{eqnarray}
Next, we will consider the application of three modified gravity models: Stelle gravity, $f(R)$ gravity, and the Brans-Dicke theory. We will investigate whether we can obtain interesting power-law solutions that satisfy either the timelike or null energy condition while potentially avoiding the big-bang singularity.
\section{Stelle Gravity}
We will start with the Stelle gravity theory that introduces higher curvature terms in an attempt to smoothen the curvature singularities at the origin. In addition to the massless graviton, it gives rise to massive spin $2$ excitations, with the action given by \cite{stelle1978classical,stelle1977renormalization,noakes1983initial} 
\begin{equation}\label{sse}
S=\int{d^4x\sqrt{-g}\left(\frac{1}{2\kappa^2}R-\alpha R_{\mu\nu}R^{\mu\nu}+\beta R^2\right)}
\end{equation}
Varying (\ref{sse}) with respect to the metric results in the following equations of motion 
\begin{equation}\label{gyt}
R_{\mu\nu}-\frac{1}{2}g_{\mu\nu}R =\kappa^2 T_{\mu\nu}+\kappa^2H_{\mu\nu},
\end{equation}
where  
\begin{multline}
H_{\mu\nu}= 2\alpha\Box R_{\mu\nu}+\left(\alpha-4\beta\right)g_{\mu\nu}\Box R+4\alpha R^{\rho\sigma}R_{\mu\rho\nu\sigma}\\-4\beta RR_{\mu\nu} -g_{\mu\nu}\left(\alpha R^{\rho\sigma}R_{\rho\sigma}-\beta R^2\right)-2\left(\alpha-2\beta\right)\nabla_{\mu}\nabla_{\nu}R.
\end{multline}

By trace-reversing equation (\ref{gyt}) we obtain
\begin{equation}\label{prwe}
R_{\mu\nu}=\kappa^2\left(T_{\mu\nu}-\frac{1}{2}g_{\mu\nu}T\right)+\kappa^2 \left(H_{\mu\nu}-\frac{1}{2}g_{\mu\nu}g^{ab}H_{ab}\right).
\end{equation}

Next, we will examine the Raychaudhuri equation to investigate the conditions for violating the convergence condition.
\subsection{Time-like Raychaudhuri Equation:}
Inserting the trace-reversed equation (\ref{prwe}) into (\ref{violation}) leads us to
\begin{multline}\label{blr}
\frac{d\theta}{d\tau}= -\frac{\theta^2}{3} -\kappa^2\left(T_{\mu\nu}u^{\mu}u^{\nu}+\frac{1}{2}T\right)\\-\kappa^2 \left(H_{\mu\nu}u^{\mu}u^{\nu}+\frac{1}{2}g^{ab}H_{ab}\right).
\end{multline}
It can be shown (see \cite{nenmeli2023stelle}) that the (00) component of the tensor $H_{\mu\nu}$ 
is given by
\begin{equation}
\kappa^2H_{00}=4\lambda\left(\dot{H}^2-2H\ddot{H}-6H^2\dot{H}\right),
\end{equation}
and the $(ii)$ components (for $i=1,2,3$) are given by
\begin{equation}
\kappa^2H_{ii} = \lambda\left(18H^2\dot{H}+9\dot{H}^2+12H\ddot{H}+2\dddot{H}\right)-2H^2-\dot{H}\,
\end{equation}
where $\lambda=(3\beta-\alpha)/\kappa^2>0$, $\alpha$ and $\beta$ are constants of the theory and $H =\frac{\dot{a}}{a}$ is the Hubble parameter. 
Including these results, the equation (\ref{blr}) becomes
\begin{multline}\label{bly}
\frac{d\theta}{d\tau}=-\kappa^2\rho_{0}\left(\frac{1+3w}{2}\right)t^{-3\alpha\left(w+1\right)}-3\alpha^2t^{-2}-6\lambda t^{-4}\\+\frac{1}{2}t^{-2\left(\alpha+1\right)}\left(1+\frac{1}{r^{2}}+\frac{1}{r^{2}\sin^{2}\theta}\right)\\-\frac{3}{2}
\lambda \ t^{-2\left(\alpha+ 2\right)}\left(1+\frac{1}{r^{2}}+\frac{1}{r^{2}\sin^{2}\theta}\right)
\end{multline} 
 The fifth term in the right hand side of the above equation (\ref{bly}) is negative and it dominates all the other terms at early times. Thus Stelle gravity leaves us with no room of possibly avoiding the big-bang singularity via the violation of the convergence theorem in the time-like case.
\subsection{The Null Raychaudhuri Equation}
The null Raychaudhuri equation on the other hand takes the form
\begin{dmath}\label{hayy}
\frac{d\hat{\theta}}{d\lambda}=-\kappa^2\rho_{0}\left(w+1\right)t^{-3\alpha\left(w+1\right)-2\alpha}
-\frac{\alpha^2}{2a_{0}^2}t^{-2(\alpha+1)}-12\lambda t^{-2(2+\alpha)}-3\lambda t^{-4(1+\alpha)}+t^{-2(1+2\alpha)}.
\end{dmath}
Since $\lambda > 0$, there is only one positive term on the right-hand side of (\ref{hayy}). However, at very early times, this positive term will not dominate the fourth term, which has the highest negative power of $t$, thereby leaving us with no chance of violating the convergence theorem. In principle, Stelle gravity does not violate singularity theorems, except in cases where $\lambda < 0$, which introduces a tachyon into the theory, and that is clearly undesirable \cite{platania2020non}.
\section{f(R) Gravity}
The second model we consider is konwn as $f(R)$ gravity, which generalizes the Einstein-Hilbert action of General Relativity by replacing the Ricci scalar $R$ with a function of the Ricci scalar \cite{buchdahl1970non}, denoted as $f(R)$. The action of $f(R)$ gravity is given by
\begin{equation}\label{sooth}
S =\frac{1}{2\kappa^2}\int{\sqrt{-g}f(R) d^4x} +\int{\sqrt{-g}\mathcal{L_{\text{m}}} d^4x}.
\end{equation}
Here, $\kappa^2 = 8\pi G/c^4$, $g$ represents the determinant of the metric tensor $g_{\mu\nu}$, and, as previously mentioned, $f(R)$ is a function of the Ricci scalar $R$. Varying (\ref{sooth}) with respect to the inverse metric tensor $g^{\mu\nu}$ results in the following field equations
\begin{equation}\label{hlodi}
f^{\prime}(R) R_{\mu\nu}-\frac{1}{2}f(R)g_{\mu\nu}+\left(\nabla_{\mu}\nabla_{\nu}-g_{\mu\nu}\Box \right)f^{\prime}(R)=\kappa^2 T_{\mu\nu},
\end{equation}
where the prime represents differentiation with respect to $R$ and $\Box = g^{\mu\nu}\nabla_{\mu}\nabla_{\nu}$. The stress energy tensor $T_{\mu \nu}$ can be obtained from matter part of the Lagrangian $\mathcal{L}_{m}$ as
\begin{equation}
T_{\mu\nu} = -\frac{2}{\sqrt{-g}} \frac{\delta\left(\sqrt{-g}\mathcal{L}_{m}\right)}{\delta g^{\mu\nu}}.
\end{equation}
When we insert the FLRW metric and the energy-momentum tensor of a perfect fluid into (\ref{hlodi}), the (00) component of the resulting tensor equation yields
\begin{eqnarray}\label{zdf}
3H^2f^{\prime} =\rho+\frac{1}{2}\left(Rf^{\prime}-f\right)-3Hf^{\prime\prime}\dot{R} 
\end{eqnarray}
In the following subsection, we will employ (\ref{zdf}) for a relatively generic choice of $f(R)$ and examine the power-law solutions that could potentially avoid the singularity.
\subsection{The $R^n$ Model}
In this paper we will consider an $f(R)$ theory consisting of the
usual scalar curvature plus a perturbative correction of higher power in the curvature as $f(R) = R[1+ (\ell_{Pl}^{2} R)^{n-1)}]$, \footnote{Note the presence of the $l_{Pl}^2$ term to ensure the correct dimensions.}The Planck length $\ell_{Pl}$ is controlling the strength of the corrections, and $n$ is a positive integer. Since we are
interested in the large curvature regime $R\gg 1/\ell^2_{Pl} \implies R\ell^2_{Pl}\gg 1$, we will consider the case where the second term in $f(R)$ dominates so that $f(R) 
\simeq R^{n}$. In this paper, this is the $f(R)$ gravity model \cite{starobinsky1980new, carloni2005cosmological,leach2006shear,de2016theoretical, carloni2006bounce} that we will investigate for singularity avoidance. 
In the following sections, we will examine the time-like and the null Raychaudhuri equations.
\subsection{Time-like Raychaudhuri Equation}
The time-like Raychaudhuri equation in $f(R) =R^n$ gravity \cite{burger2018towards} is given by
\begin{equation}\label{afl}
\frac{d\theta}{d\tau} =-\frac{\theta^2}{3}-\frac{\kappa^2T_{\mu\nu}u^{\mu}u^{\nu}}{2n\ell_{Pl}^{2n-2}R^{n-1}}
+\frac{3}{n}\left(\frac{\ddot{a}}{a}+\frac{\dot{a}^2}{a^2}\right).
\end{equation}
For Einstein gravity ($n=1$), the second term in the above equation will read
\begin{equation}
-\frac{\kappa^2T_{\mu\nu}u^{\mu}u^{\nu}}{2n\ell_{Pl}^{2n-2}R^{n-1}}=-\frac{\kappa^2}{2}T_{\mu\nu}u^{\mu}u^{\nu}, 
\end{equation}
which is not supressed by the large curvature term, unlike the $R^n$ case, where the stress energy term becomes subleading due to the presence of the $R^{n-1}$ term in the denominator.

The remaining sum (without the second term) of (\ref{afl}) in Einstein's gravity reduces to
\begin{equation}
-\frac{\theta^2}{3}+\frac{3}{n}\left(\frac{\ddot{a}}{a}+3\frac{\dot{a}^2}{a^2}\right)=3\frac{\ddot{a}}{a}
\end{equation}
which is always negative since 3$\frac{\ddot{a}}{a}<0$ for standard matter (except the cosmological constant) hence ruling out a possibility of avoiding the big bang singularity. When $n\neq 1$ equation (\ref{afl}) in the large curvature regime $\ell^2_{Pl}R\gg 1$ reduces to
\begin{dmath}\label{arw}
\frac{d\theta}{d\tau}=-3\alpha^2t^{-2}+\frac{3}{n}\left(2\alpha^2-\alpha\right)t^{-2}.
\end{dmath}
This gives rise to the necessary positive term on the right-hand side of the Raychaudhuri equation. Therefore, extending the model from Einstein gravity to $f(R)$ gravity accomplishes two things:
Firstly, it makes the stress-energy term, which appears as negative, subleading, and secondly, it introduces the necessary positive term, which is not present in the Einstein case.

Now, to violate the convergence condition the sum of the coefficients of $t$ in (\ref{arw}) must be positive:
\begin{equation}\label{qaww}
\frac{3 \alpha (2 \alpha-1)}{n}-3\alpha^2>0.
\end{equation}
Thus the condition simplifies as
\begin{equation}\label{fqa}
3\alpha^2\left(\frac{2}{n}-1\right)-\frac{3\alpha}{n}>0.
\end{equation}
\begin{figure}
 \centering
 \captionsetup{justification=raggedright,
singlelinecheck=false}
 \includegraphics[width=0.9\linewidth]{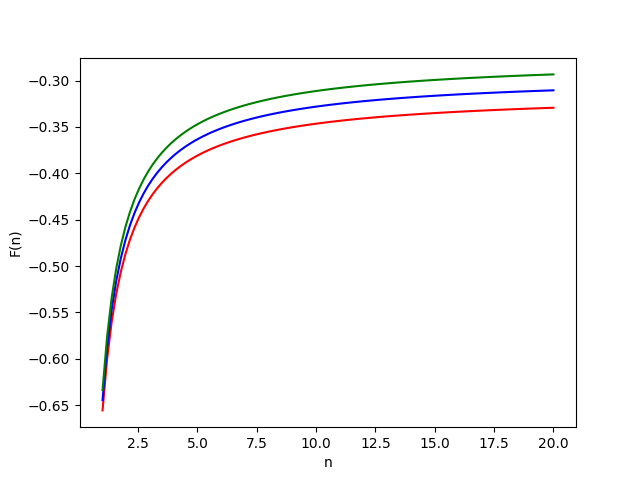} 
\caption{Displays the left hand side of the inequality Eq.(\ref{fqa}) for $\alpha=0.323 $ (red), $\alpha=0.313$ (blue) and $\alpha=0.303$ (green). }\label{fig:gde}
\end{figure}
\begin{figure}
 \centering
 \captionsetup{justification=raggedright,
singlelinecheck=false}
 \includegraphics[width=0.9\linewidth]{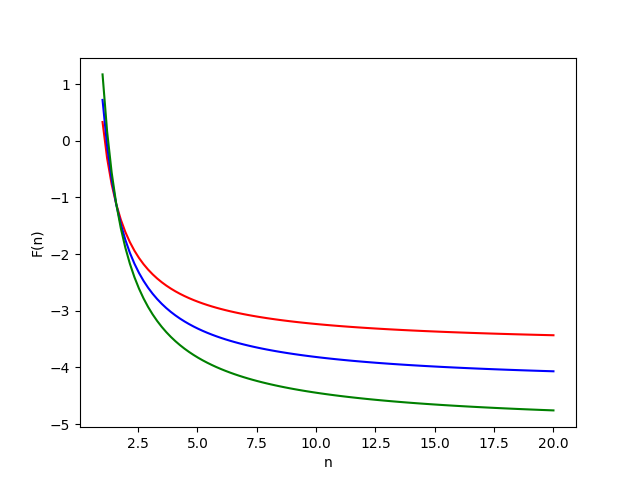} 
\caption{Displays the left hand side of the inequality Eq.(\ref{fqa}) for $\alpha=1.1$ (red), $\alpha=1.2$ (blue) and $\alpha=1.3$ (green). }\label{fig:gle}
\end{figure}
The first observation to make is that for $n=2$, the left-hand side of the inequality (\ref{fqa}) is not defined. Therefore, we need to investigate the inequality for $n>2$. Next, we will plot the left-hand side of the inequality (\ref{fqa}) for ekpyrosis ($\alpha<\frac{1}{3}$) (see Fig. \ref{fig:gde}) and for the accelerating case ($\alpha>1$) (Fig. \ref{fig:gle}). Fig. \ref{fig:gde} clearly shows that there is no interval of values of $n>2$ for which the inequality in (\ref{fqa}) is satisfied. Thus, ekpyrosis is ruled out. However, for the accelerating case, there is a small interval of $n$ for which the inequality holds. This implies that in the time-like case, we only obtain the accelerating solution. Next we will consider the null case for power-law solutions. 
\subsection{The Null Raychaudhuri Equation}
\begin{figure}
 \centering
 \captionsetup{justification=raggedright,
singlelinecheck=false}
 \includegraphics[width=0.9\linewidth]{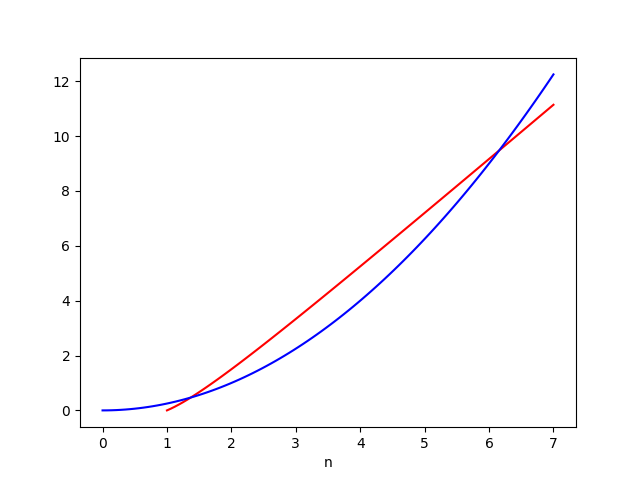} 
\caption{Displays the left hand side (red) and the right hand side (blue) of the inequality Eq.(\ref{ast}).}\label{fig:glo2}
\end{figure}

The null Raychaudhuri equation in $f(R) = R^n$ gravity takes the following form \cite{burger2018towards},
\begin{equation}\label{awl}
\frac{d\hat{\theta}}{d\lambda}=-\frac{\hat{\theta}^2}{2}-\frac{\kappa^2T_{\mu\nu}n^{\mu}n^{\nu}}{2n\ell_{Pl}^{2n-2}R^{n-1}}+\frac{a^{-2}}{n R^{n-1}}\partial_{t}R^{n-1}.
\end{equation}
We will see that, just like the time like case, extending the model from Einstein gravity to $R^n$ model makes the stress-energy term, subleading, and will introduce the necessary positive term.
To see this explicitly, consider $n=1$, then the stress energy term simplifies as follows
\begin{equation}
-\frac{\kappa^2T_{\mu\nu}n^{\mu}n^{\nu}}{2n\ell_{Pl}^{2\left(n-1\right)}R^{n-1}} \simeq -\frac{\kappa^2}{2}T_{\mu\nu}n^{\mu}n^{\nu},
\end{equation}
which is clearly not suppressed by higher curvature term. On top of that the necessary positive term
\begin{equation}
+\frac{1}{R^{n-1}}\left(n^t\right)^2\partial^2_{t}R^{n-1}=0 
\end{equation}
will not exist. 
The expression (\ref{awl}) above can thus be expressed as
\begin{dmath}
\frac{d\hat{\theta}}{d\lambda} =-\frac{1}{2}\alpha^2t^{-2\alpha-2}+\frac{\left(2-2n\right)\left(1-2n\right)}{n}t^{-2\alpha-2}. 
\end{dmath}
Just as in the time-like case, the time dependence on the right-hand side of the above equation is identical for all terms. Ignoring the Planck-suppressed term, the condition for violating the focusing theorem simplifies to:
\begin{equation}\label{ast}
\frac{\left(1-n\right)\left(1-2n\right)}{n}> \left(\frac{\alpha} {2}\right)^2.
\end{equation}
As expected for Einstein gravity ($n=1$), the above inequality will not be satisfied. However, for any $n$ between $(1.21, 6.16)$, we obtain an accelerating solution (Fig. \ref{fig:glo2}). On the other hand, we get ekpyrosis for any $n>1$.

\section{Brans-Dicke theory}
The last model of interest is the Brans-Dicke theory of gravitation, which incorporates an additional long-range scalar field alongside the metric and is considered a viable alternative to General Relativity. The action for the Brans-Dicke theory in the Jordan frame is given by \cite{brans1961mach, weinberg2004gravitation}
\begin{multline}\label{trt}
 S =\frac{1}{16\pi}\int{d^4x\sqrt{-g}\left(\phi R-\frac{\omega}{\phi}\nabla_{\mu}\phi\nabla^{\mu}\phi\right)}  
 \\+\int{d^4x\sqrt{-g}\mathcal{L}_{M}}
\end{multline}
where $\mathcal{L}_{M}$ is the Lagrangian density of matter, $\phi$ is the scalar field of the theory, $g$ is the determinant of the metric tensor $g_{\mu\nu}$ and $\omega$ is the Brans-Dicke coupling constant. By varying (\ref{trt}) with respect to the metric tensor, we get the field equations of motion
\begin{dmath}\label{prt}
 R_{\mu\nu}  -\frac{1}{2}R g_{\mu\nu} = \frac{8\pi}{\phi} T_{\mu\nu}+\frac{1}{\phi}\left(\nabla_{\mu}\nabla_{\nu}\phi-g_{\mu\nu}\Box\phi\right)+\frac{\omega}{\phi^2}\left(\nabla_{\mu}\phi\nabla_{\nu}\phi-\frac{1}{2}g_{\mu\nu}\nabla_{c}\phi\nabla^{c}\phi\right).
\end{dmath}
In addition, the variation of the action with respect to the scalar field results in the equation of motion for $\phi$
\begin{equation}\label{prot}
 \Box\phi = \frac{8\pi}{3+2\omega}T.
\end{equation}

In Brans-Dicke theory, the scalar field $\phi$ is related to the gravitational constant G by:
\begin{equation}\label{aust}
\text{G}=\frac{1}{\phi},
\end{equation}
when we compare the first term on the right hand side of (\ref{prt}) to the source term in the usual Einstein field equations of GR. The relation between $\phi$ and $\text{G}$ in (\ref{aust}) indicates that $\text{G}$ varies with time. The coupling constant $\omega$ is a tunable parameter that can be adjusted to fit the data at hand.
\subsection{Dynamics from the Friedmann Equations}
In this section we will consider power-law solution for the scale factor as 
\begin{equation}
a(t) \propto t^{\alpha}. 
\end{equation}
The (00) component of the Einstein-Brans-Dicke equation (\ref{prt}) 
for perfect fluid takes the form: 
\begin{equation}\label{ytr}
 3\left(\frac{\dot{a}}{a}\right)^2 = \frac{\rho}{\phi}-3\frac{\dot{a}}{a}\frac{\dot{\phi}}{\phi}+\frac{\omega}{2}\left(\frac{\dot{\phi}}{\phi}\right)^2 
\end{equation}
One can re-express (\ref{ytr}) as 
\begin{equation}\label{mpoo}
\frac{\omega}{2}=3H^2\left(\frac{\phi}{\dot{\phi}}\right)^2+3H\frac{\phi}{\dot{\phi}}-\frac{\rho\phi}{\dot{\phi}^2}.
\end{equation}
The RHS of this equation is fixed since the Brans-Dicke parameter $\omega$ is a constant. The first and second terms in equation (\ref{mpoo}) are constants since both $a(t)$ and $\phi(t)$ are power-law and the time dependence exactly cancels out. Therefore, the last term should be a constant. This implies that
\begin{equation}\label{olala}
\frac{\rho\phi}{\dot{\phi}^2} = \text{constant}   
\end{equation}
which reduces to 
\begin{equation}\label{yut}
\frac{\dot{\phi}^2}{\phi} \propto \rho.
\end{equation}
Inserting the power law solutions of $a(t) \sim t^{\alpha}$, $\phi(t) = t^{\gamma}$ and $\rho\propto a^{-3\left(w+1\right)}$ into (\ref{yut}) results in
\begin{equation}
t^{\gamma-2} \propto t^{-3\alpha\left(1+w\right)}.
\end{equation}
We thus get the following relation between the exponents:
\begin{equation} \label{alpha}
\alpha =\frac{2-\gamma}{2\epsilon}.
\end{equation}
Note that when expressed in terms of $\epsilon$ the strong and null energy conditions read $\epsilon \geq 1 $ and $\epsilon \geq 0$ respectively. From Eq. \ref{alpha}, we also observe that the largest value of $\gamma$ approaches 2 to ensure that the exponent $\alpha$ is positive when we demand the strong energy condition. The same condition applies to the null energy case, except for $\epsilon=0$, for which $\alpha$ is not defined. 

\subsection{Time-like Raychaudhuri Equation}
The curvature term of the Brans-Dicke theory (when $k=0$) is given by  \cite{choudhury2021raychaudhuri}
\begin{equation}\label{bty}
 R_{\mu\nu}u^{\mu}u^{\nu} = \frac{\omega\left(3p-\rho\right)}{\left(2\omega+3\right)\phi}+3H^2+\frac{\omega}{2}\left(\frac{\dot{\phi}}{\phi}\right)^2.
\end{equation}
Inserting the equation of state $p=w\rho$ and $\epsilon$ into (\ref{bty}) results in
\begin{equation}\label{xvt}
R_{\mu\nu}u^{\mu}u^{\nu} = \left ( \frac{\omega\left(2\epsilon-4\right)}{2\omega+3}+3\alpha^2+\frac{\gamma^2 \omega}{2} \right ) \frac{1}{t^2}.
\end{equation}
The time-like Raychaudhuri equation in FLRW spacetime thus becomes
\begin{equation}\label{chau}
\frac{d\theta}{d\tau}= -\left ( \frac{\omega\left(2\epsilon-4\right)}{2\omega+3}+6\alpha^2+\frac{\gamma^2 \omega}{2} \right ) \frac{1}{t^2}.
\end{equation}
We immediately notice that the LHS of (\ref{chau}) is positive when
\begin{equation}\label{zow}
\frac{\omega\left(2\epsilon-4\right)}{2\omega+3}+6\alpha^2+\frac{\gamma^2\omega}{2}<0.
\end{equation}
We can rewrite it as follows:
\begin{equation}\label{zow2}
\frac{\omega\left(2\epsilon-4\right)}{2\omega+3}+\frac{\gamma^2\omega}{2}<-6\alpha^2.
\end{equation}
Since $6\alpha^2\geq 0$ we can first look into the weaker version of (\ref{zow2}) as follows:
\begin{equation}\label{zaw}
\frac{\omega\left(\left(2\epsilon-4\right)+\frac{\gamma^2}{2}\left(2\omega+3\right)\right)}{\left(2\omega+3\right)}<0.
\end{equation}

In the following analysis, we will initially determine a permissible set of values for $\gamma$ based on this condition and subsequently assess whether any of these allowed values align with the complete inequality in (\ref{zow2}). The inequality (\ref{zaw}) is satisfied in the following ways: \\

\noindent
{\bf Case 1:}
\begin{eqnarray} \label{casr}
-\frac{3}{2}<\omega &<0  \\
2\epsilon-4+\frac{\gamma^2}{2}\left(2\omega+3\right)&>0. \label{casr2}
\end{eqnarray}

\noindent
{\bf Case 2:}
\begin{eqnarray}\label{awq}
\omega &<-\frac{3}{2}\\
2\epsilon-4+\frac{\gamma^2}{2}\left(2\omega+3\right)&< 0. \label{awq2}
\end{eqnarray}

\noindent
{\bf Case 3:} 
\begin{eqnarray}\label{awq3a}
\omega &> 0 \\
2\epsilon-4+\frac{\gamma^2}{2}\left(2\omega+3\right)&< 0. \label{awq3b}
\end{eqnarray}
Note that Eqs. (\ref{awq2}) and (\ref{awq3b}) are identical. In order to violate the convergence condition, $\omega$ needs to obey either case 1, case 2, or case 3. The bounds on $\omega$ in case 1 and case 2 were discussed in \cite{banerjee2001cosmic}, where the latter case was shown to model accelerated expansion in Brans-Dicke theory and the former was mentioned to be consistent with a radiation-dominated epoch of the theory. Note that when $\gamma=0$, $\phi$ becomes a constant, and the power-law solution takes the form $a(t)\propto t^{\frac{1}{\epsilon}}$, thus recovering GR. \\
\begin{figure}
 \centering
 \captionsetup{justification=raggedright,
singlelinecheck=false}
 \includegraphics[width=0.9\linewidth]{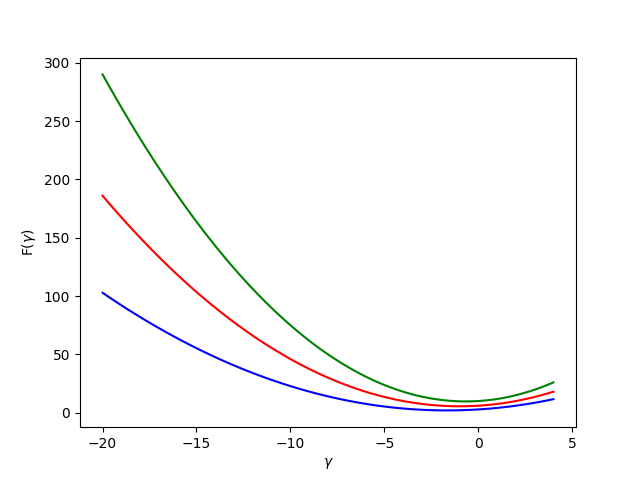} 
\caption{Displays the left hand side of the inequality Eq.(\ref{full0}) for a fixed value of $\omega = -3.5/2$ (consistent with Eq.(\ref{awq}) and for various values of $\alpha = 1.2, 2, 3$ that is larger than 1. }\label{fig:gamma}
\end{figure}

\noindent
\textbf{Accelerating Power-law Solution:}
First, we will explore the feasibility of an accelerating power-law solution that has the potential to bypass the singularity theorems and thereby provide a potential 
avenue of resolving the cosmological singularity. 
Specifically, we explore if there can be a repulsive contribution to the RHS of the Raychaudhuri equation. For an accelerating solution, we have the scale factor as $\alpha > 1$. We will investigate the above mentioned cases below. \\

\begin{figure}
 \centering
 \captionsetup{justification=raggedright,
singlelinecheck=false} 
\includegraphics[width=0.9\linewidth]{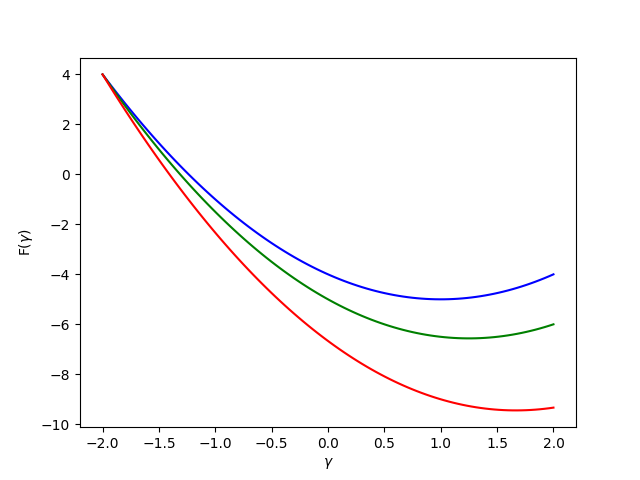} 
\caption{Displays the left hand side of the inequality Eq.(\ref{xcv}) for various values of $-\frac{3}{2}<\omega<0$ consistent with \textbf{Case 1}. The blue graph corresponds to $\omega=-1$, green $\omega=-1.1$ and red $\omega=-1.2$.}
\label{fig:ekpy5}
\end{figure}

\noindent
\textbf{Case 1}: When we substitute $\alpha=1$ into Eq.(\ref{casr2}) we get
\begin{equation}\label{xcv}
\gamma^2-\frac{2\omega}{2\omega+3}-\frac{4}{2\omega+3}<0\,.
\end{equation}
In Fig. \ref{fig:ekpy5}, we observe an interval of values for $\gamma$ in which the left-hand side of the inequality (\ref{xcv}) is satisfied. This imposes a constraint on the type of $\gamma$ required to achieve accelerated expansion in Brans-Dicke theory.\\

\noindent
{\bf Case 2:} By using the expression for $\alpha$ in (\ref{awq2}) we get
\begin{figure}
 \centering
\captionsetup{justification=raggedright,
singlelinecheck=false}
   \includegraphics[width=0.9\linewidth]{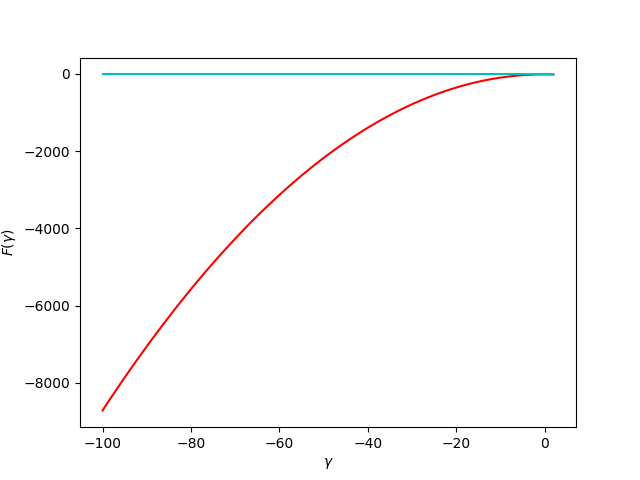} 
      \caption{The left hand side (red) and right hand side (cyan) of Eq.(\ref{zow2}) for the values of $\gamma$ that is consistent with Eq.(\ref{full0}) for a fixed value of $\omega = -3.5/2$ that satisfies the inequality $\omega <-3/2$. } \label{fig:gamma2}
\end{figure}
\begin{figure}
 \centering
\captionsetup{justification=raggedright,
singlelinecheck=false}
\includegraphics[width=0.9\linewidth]{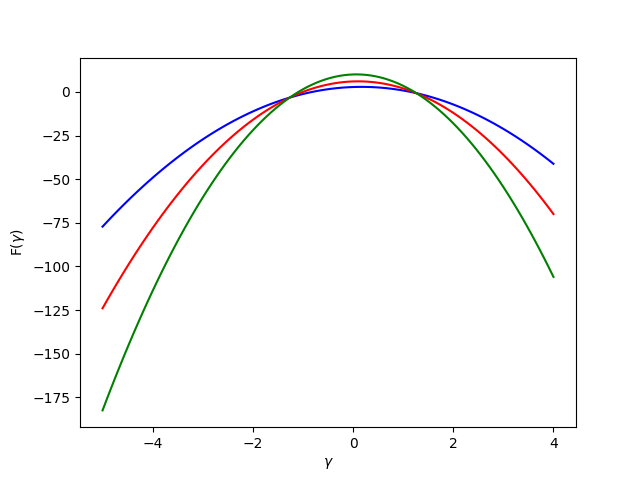} 
\caption{Displays the left hand side of the inequality Eq.(\ref{full0}) for a fixed value of $\omega = 1$ and for various values of $\alpha = 1.1, 2, 3$ that is larger than 1. }\label{fig:case3}
\end{figure}

\begin{equation}
\gamma > 2- \alpha \left(4-\frac{\gamma^2}{2}\left(2\omega+3\right) \right)
\end{equation}
which can be reduced to
\begin{equation} \label{full0}
\gamma - 2 + \alpha \left(4-\frac{\gamma^2}{2}\left(2\omega+3\right) \right) >0
\end{equation}
\begin{figure}
 \centering
\captionsetup{justification=raggedright,
singlelinecheck=false}
   \includegraphics[width=0.9\linewidth]{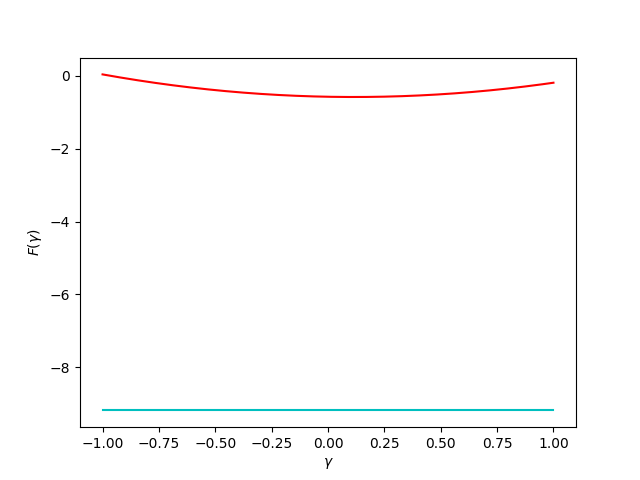} 
      \caption{The left hand side (red) and right hand side (cyan) of Eq.(\ref{zow2}) for $-1\leq\gamma\leq1$ and a fixed value of $\omega = 1$ that satisfies the inequality $\omega >0$. } \label{fig:case3final}
\end{figure}
In figure \ref{fig:gamma} we display the LHS of the inequality (\ref{full0}) for a fixed value of $\omega = -3.5/2$, which is consistent with the other condition \ref{awq}) and for various values of $\alpha = 1.2, 2, 3$ which are larger than 1. We see that there is an interval of values of $\gamma$ for which the expression in (\ref{full0} )is positive, hence satisfies the inequality. 

Next we need to see if the allowed $\gamma$ can satisfy (\ref{zow2}). In Fig \ref{fig:gamma2} we show the LHS (red) and RHS (cyan) of (\ref{zow2}) for the values of $\gamma$, which is consistent with (\ref{full0}) for a fixed value of $\omega = -3.5/2$. We clearly see that the LHS is less than the RHS. Therefore, we can easily get accelerating solutions. \\

\noindent
{\bf Case 3:} Just like in the previous case, when we use (\ref{alpha}), we arrive at the same condition as in (\ref{full0}) and find that there is no interval of values for $\gamma$ in which the expression in (\ref{full0}) is positive; thus, the inequality is not satisfied. In Fig. \ref{fig:case3final}, we plotted the LHS (red) and RHS (cyan) of (\ref{zow2}) in the interval $-1\leq\gamma\leq 1$, with a fixed value of $\omega = 1$. It is evident that the RHS is greater than the LHS, indicating the absence of accelerating solutions.
\vskip 2mm

\noindent
\textbf{Ekpyrosis in Brans-Dicke Theory:}
The ekpyrotic or cyclic cosmology stands out as one of the most extensively discussed 
alternative to the theory of cosmic inflation \cite{guth1981inflationary,linde1982new}. According to this idea, the ongoing expansion of our universe is just one occurrence in an infinite series of cycles. Each cycle commences with a big bang phase, undergoes a gradual acceleration, transitions into a slow contraction phase, and concludes with a big crunch. In this study, we have also investigated these ekpyrotic solutions. Searching for ekpyrotic solutions is mathematically very similar to the search for accelerating power-law solutions. To get ekpyrosis we need to satisfy the following condition
coming from Eq.(\ref{alpha}),
\begin{equation}\label{hhr}
3(2- \gamma) < 2 \epsilon.
\end{equation}

\begin{figure}
 \centering
\captionsetup{justification=raggedright,
singlelinecheck=false}
\includegraphics[width=0.9\linewidth]{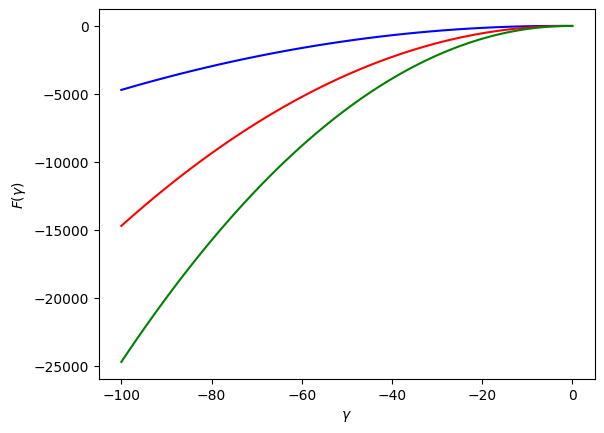} 
\caption{Displays the left hand side of the inequality Eq.(\ref{xae}) for various values of $\omega < -3/2$ (consistent with \ref{awq}).}\label{fig:ekpy}
\end{figure}

\begin{figure}
\centering
\captionsetup{justification=raggedright,
singlelinecheck=false}

\includegraphics[width=0.9\linewidth]{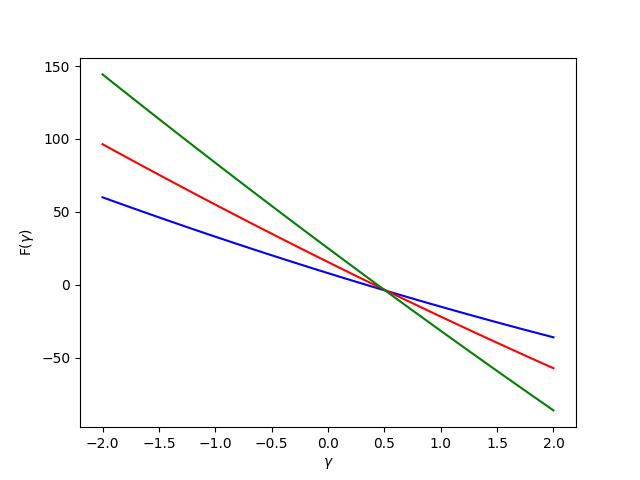} 
\caption{Displays the left hand side of the inequality Eq.(\ref{xas}) when $\omega=-1$ and $\alpha=0.25$ (blue), $\omega=-1.1$ and $\alpha=0.125$ (red) and $\omega=-1.2$ and $\alpha=0.0625$ (green) which are consistent with $-\frac{3}{2}<\omega<0$ in \textbf{Case 1}.}\label{ekpy4}
\end{figure}

\noindent
\textbf{Case 1}: By inserting (\ref{hhr}) into (\ref{casr2}) we get:
\begin{equation}\label{xas}
\gamma^2-\frac{2\gamma}{\alpha\left(2\omega+3\right)}+\frac{4}{\alpha\left(2\omega+3\right)}-\frac{8}{2\omega+3}<0.
\end{equation}
We proceed with a similar analysis as in the accelerating case. Fig. \ref{ekpy4} illustrates that there exists an interval for $\gamma$ in which the left-hand side of (\ref{xas}) is negative. The figure is plotted for various values of $\alpha$ and $\omega$ specified in the caption. Therefore, we can easily observe ekpyrosis.\\

\noindent
\textbf{Case 2}: Comparing (\ref{hhr}) with the second convergence violating bounds in (\ref{awq2}), we get the following inequality,
\begin{equation}
3(2-\gamma) < 4-\frac{\gamma^2}{2}\left(2\omega+3\right),
\end{equation}
which can be rewritten as:
\begin{equation}\label{xae}
\frac{\gamma^2}{2} (2 \omega +3) -3 \gamma +2 < 0.
\end{equation}
As in the accelerating case, the largest possible value of $\gamma$ is 2. Now we will plot (Fig. \ref{fig:ekpy}) the LHS of the inequality \ref{xae} for $\omega<-3/2$ to see if there is any $\gamma$ that satisfies (\ref{xae}). 
\begin{figure}
 \centering
\captionsetup{justification=raggedright,
singlelinecheck=false}
 
   \includegraphics[width=0.9\linewidth]{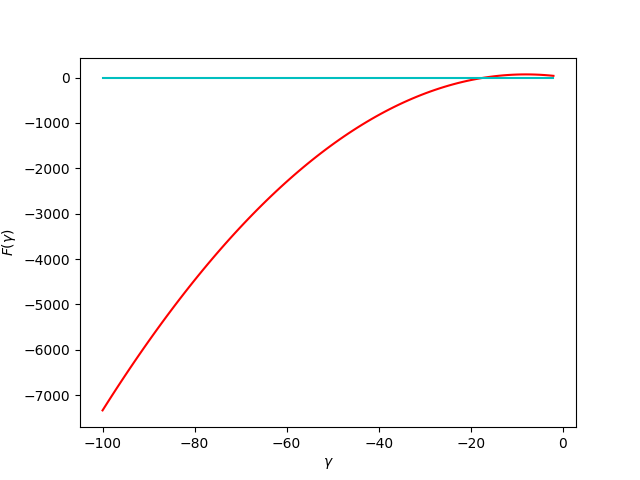} 
      \caption{The left hand side (red) and right hand side (cyan) of Eq.(\ref{zow2}) for the values of $\gamma$ that is consistent with Eq.(\ref{full0}) for a fixed value of $\omega = -3.5/2$ and $\alpha =1/4$. } \label{fig:ekpy2}
\end{figure}
It is clear from Fig.\ref{fig:ekpy2} that we can easily get ekpyrotic solutions.\\

\begin{figure}
 \centering
\captionsetup{justification=raggedright,
singlelinecheck=false}
 
\includegraphics[width=0.9\linewidth]{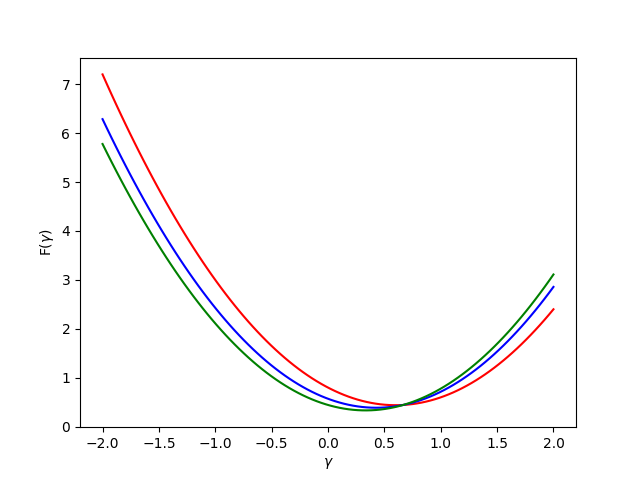} 
\caption{Displays the left hand side of the inequality Eq.(\ref{xae}) when $\omega=1$ (red), $\omega=2$ (blue) and $\omega=3$ (green) which are consistent with $\omega>0$ in \textbf{Case 3}.}\label{fig:ekpy3}
\end{figure}

\noindent
\textbf{Case 3}:
For illustration, let us define 
\begin{equation}
F(\gamma)\equiv \gamma^2-\frac{6\gamma}{2\omega+3}+\frac{4}{2\omega+3}. 
\end{equation}
Then, equation \ref{xae} reduces to $F(\gamma)<0$. A plot of $F(\gamma)$ (see Fig. \ref{fig:ekpy3}) when $\omega=1,2,3$ and $-2\leq\gamma\leq2$ reveals that $F(\gamma)$ cannot be negative. This implies that for values of $\gamma$ in the interval $-2\leq \gamma\leq 2$, equation (\ref{xae}) cannot hold, and thus, there is no ekpyrosis for this case.

\subsection{The Null Raychaudhuri Equation}
Finally we will investigate the Null Raychaudhuri equation in Brans-Dicke theory. The equation takes the form:
\begin{dmath}\label{afg}
\frac{d\hat{\theta}}{d\lambda} =-8\pi\left(w+1\right)
t^{-3\alpha\left(w+1\right)-\gamma-2\alpha}\\-\left[\gamma^2\omega+\gamma\left(\gamma-1\right)\right]t^{-2(1+\gamma)}
-\frac{\alpha^2}{a_{0}^2}t^{-2(\alpha+1)}.
\end{dmath}
To violate the convergence condition we need a dominating positive term in the right hand side of the above expression. Therefore, first we need to impose the following condition
\begin{equation}\label{alt}
\gamma^2\omega+\gamma\left(\gamma-1\right)<0. 
\end{equation}
We notice that by using (\ref{alpha}), the time exponent of the first term on the right hand side becomes
\begin{equation}
3\alpha(w+1)+\gamma+2\alpha= 2(\alpha+1),
\end{equation}
which is the same time exponent as the third term. 
Therefore, we can violate the convergence condition in two ways. The first one occurs when 
\begin{equation}\label{ula}
\alpha < \gamma.
\end{equation}

\begin{figure}
\centering
\captionsetup{justification=raggedright,
singlelinecheck=false}
\includegraphics[width=0.9\linewidth]{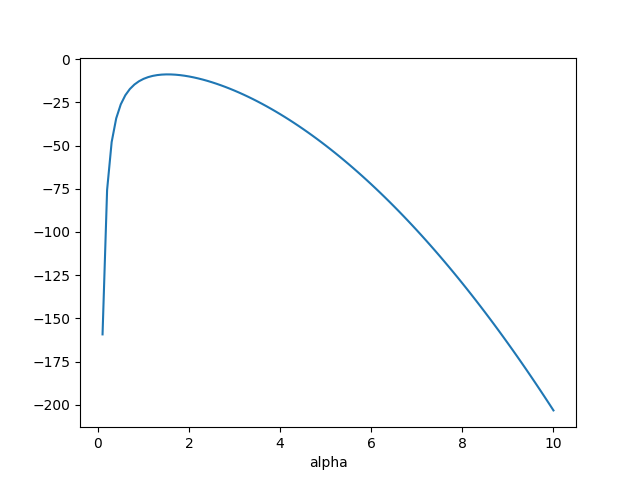} 
\caption{Displays the left hand side of the inequality Eq.(\ref{awt}) when $\omega<-1.5$ and $0<\alpha<10$ - which covers the bounds for both ekpyrosis and the accelerating solutions.}
\label{ekpw}
\end{figure}
By plotting the left hand side of this inequality in Fig. \ref{ekpw} we see that neither the ekpyrotic nor the accelerating solutions obey Eq. (\ref{awt}). Therefore, none of the power-law solutions in the null case can violate the convergence condition. For this case the powers of $t$ would not be able to reproduce (\ref{alpha}) and so we discard this option as a resolution to violate the convergence condition. This leave us to the other case that appears when $\gamma = \alpha$. For this case to violate the convergence condition we need to compare the coefficients and we get the following condition
\begin{equation}\label{ruo}
\lvert \gamma^2\left(\omega+1\right) -\gamma\rvert  >\left(8 \pi (w+1)+\frac{\alpha^2}{a_0^2} \right).
\end{equation}
Substituting $\alpha=\gamma$ in the above we get a condition in terms of $\alpha$ as
\begin{equation}\label{awt}
\alpha^2\omega-\alpha-\frac{8\pi}{3}\left(\frac{2-\alpha}{\alpha}\right)>0.
\end{equation}


\section{Conclusions}
\begin{center}
\captionof{table}{Table showing convergence ($\times$) and
divergence (check) for the various models and the 
time-like and null Raychaudhuri equations.}
\begin{tabular}{|c|c|c|c|}
\hline
{} Model&Power-law& Avoid Singularity & Avoid Singularity\\
         &Solution& Time-like RE& Null RE\\     
\hline  
Stelle &Ekpyrosis & $\times$ & $\times$ \\ \hline                  
        \cline{2-2}
        & Accelerating& $\times$ & $\times$ \\
\hline  
$R^n$&Ekpyrosis &$\times$&\checkmark\\ \hline                  
        \cline{2-2}
        & Accelerating&\checkmark& \checkmark\\
\hline
B-D &Ekpyrosis &\checkmark&$\times$\\ \hline                  
        \cline{2-2}
        & Accelerating&\checkmark&$\times$\\
\hline
\end{tabular}
\end{center}

In this article, we explored some of the promising modifications of GR which have attempted to get rid of short distance singularities, the latter being its generic feature, as well as address other outstanding issues. Considering the power-law solutions for the scale factor, our strategy has been to examine the time-like and null Raychaudhuri equations for these theories, since the prediction of the equation that neighbouring geodesics converge to conjugate points in a finite proper time, is the strongest indicator, and in fact a necessary condition for the existence of singularities. 
Therefore, any dominant repulsive term on the right hand side of the Raychaudhuri equation in the early universe would be a strong indication of singularity avoidance, which merits further study. 

Our main results are summarized in the accompanying table. As can be seen, Stelle gravity does not offer a viable way to avoid a singularity.
On the other hand, for $f(R)$ gravity, the
accelerating solutions seems more natural. 
Finally, for Brans-Dicke theory, both 
solutions are viable only for the strong
energy condition. We may note that in these theories, ekpyrosis still remains a potential candidate for the resolution of singularities. 
While this does not prove beyond doubt that 
power law solutions are feasible options to consider in the neighbourhood of the putative singularity, it certainly merits further investigation. We hope to report on this elsewhere. 
\begin{acknowledgments}
The authors thank P. Dunsby and M. Fridman for 
their useful comments which helped improve the manuscript. 
This work was supported by the Natural Sciences and Engineering
Research Council of Canada. SSH, and ST are supported in part by the “New insights into Astrophysics and Cosmology with theoretical models confronting observational data” program from the National Institute for Theoretical and Computational Sciences of South Africa (NITheCS).

\end{acknowledgments}

\newpage
\bibliography{main_final}
\end{document}